\documentclass{emulateapj}
\pdfoutput=1
\submitted{Accepted by ApJL}

\newcommand{\src}{G1.9+0.3}
\newcommand{\tbn}{$\theta_{\rm Bn}$}
\newcommand{\roll}{$\nu_{\rm roll}$}

\catcode`\@=11
\newcommand{\gapprox}{\mathrel{\mathpalette\@versim>}}
\newcommand{\lapprox}{\mathrel{\mathpalette\@versim<}}
\newcommand{\propapprox}{\mathrel{\mathpalette\@versim\propto}}
\newcommand{\@versim}[2]
  {\lower3.1truept\vbox{\baselineskip0pt\lineskip0.5truept
\ialign{$\m@th#1\hfil##\hfil$\crcr#2\crcr\sim\crcr}}}
\catcode`\@=12

\shorttitle{YOUNGEST GALACTIC SNR G1.9+0.3}

\begin{document}

\title{X-ray Spectral Variations in the 
Youngest Galactic Supernova Remnant  G1.9+0.3}


\author{Stephen P. Reynolds,\altaffilmark{1}
Kazimierz J. Borkowski,\altaffilmark{1}
David A. Green,\altaffilmark{2}
Una Hwang,\altaffilmark{3}
Ilana Harrus,\altaffilmark{3}
\& Robert Petre \altaffilmark{3}}

\altaffiltext{1}{Department of Physics, North Carolina State University,
  Raleigh NC 27695-8202; stephen\_reynolds@ncsu.edu} 
\altaffiltext{2} {Cavendish Laboratory; 19 J.J. Thomson Ave., 
Cambridge CB3 0HE, UK}
\altaffiltext{3}{NASA/GSFC, Code 660, Greenbelt, MD 20771}

\begin{abstract}

The discovery of the youngest Galactic supernova remnant (SNR)
G1.9+0.3 has allowed a look at a stage of SNR evolution never before
observed.  We analyze the 50 ks Chandra observation with particular
regard to spectral variations.  The very high column density ($N_H
\sim 6 \times 10^{22}$ cm$^{-2}$) implies that dust scattering is
important, and we use a simple scattering model in our spectral
analysis.  The integrated X-ray spectrum of \src\ is well described by
synchrotron emission from a power-law electron distribution with an
exponential cutoff.  Using our measured radio flux and including
scattering effects, we find a rolloff frequency of $5.4 (3.0, 10.2)
\times 10^{17}$ Hz ($h \nu_{\rm roll} = 2.2$ keV).  Including
scattering in a two-region model gives lower values of \roll\ by over
a factor of 2.  Dividing \src\ into six regions, we find a systematic
pattern in which spectra are hardest (highest \roll) in the bright SE
and NW limbs of the shell.  They steepen as one moves around the shell
or into the interior.  The extensions beyond the bright parts of the
shell have the hardest spectra of all.  We interpret the results in
terms of dependence of shock acceleration properties on the obliquity
angle $\theta_{\rm Bn}$ between the shock velocity and a fairly
uniform upstream magnetic field.  This interpretation probably
requires a Type Ia event.  If electron acceleration is limited by
synchrotron losses, the spectral variations require
obliquity-dependence of the acceleration rate independent of the
magnetic-field strength.

\end{abstract}

\keywords{
supernova remnants ---
ISM: individual (G1.9+0.3) ---
X-rays: ISM
}

\section{Introduction}
\label{intro}

\begin{figure}
\vspace{-1.7truein}
\epsscale{1.3}
\hspace{-0.2truein}
\plotone{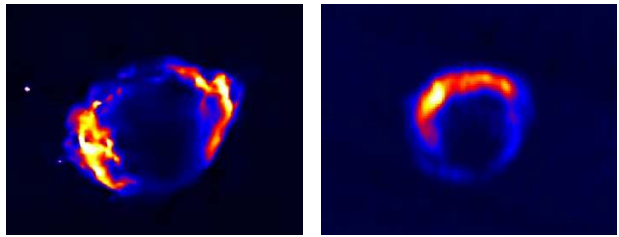}
\vspace{-1.8truein}
\caption{Left: {\sl Chandra} image of \src, platelet smoothed (Willett
2007).  Colors are intensities only, between 1.5 and 6 keV.  
Image size $136'' \times 185''$.  Right:
2008 radio image of \src\ (Green et al.~2008): VLA at 4.9 GHz.
Resolution $10'' \times 4''$.
\label{xrim}}
\end{figure}

The supernova remnant (SNR) G1.9+0.3 has recently been shown to have
expanded by about 16\% between 1985 and 2007, implying an age of order
100 years -- the youngest supernova remnant in the Galaxy
\citep[][hereafter Paper I]{reynolds08b}.  The expansion was confirmed
with new VLA observations from March 2008 \citep{green08}.  The X-ray
and current radio images are shown in Fig.~\ref{xrim}.  The X-ray
spectrum is featureless and well described by the loss-steepened tail
of the synchrotron spectrum inferred from radio frequencies.  No
thermal X-ray emission is apparent.  \src\ provides a unique
opportunity to study a SNR at a stage never before observed, and to
learn about the physics of shock acceleration in faster shocks than
seen in any SNR ($v_{\rm s} \sim 14,000$ km s$^{-1}$; Paper I).

A simple model of a power-law electron distribution with an
exponential cutoff at energy $E_{\rm max}$ producing synchrotron 
radiation (XSPEC model {\tt srcut}) has proved to be a useful tool in
understanding X-ray synchrotron spectra in those dozen or so SNRs in
which the phenomenon is observed \citep{reynolds08a}.  The synchrotron
spectrum cuts off more slowly than exponential, roughly as $S_X
\propto \exp[-(\nu/\nu_{\rm roll})^{1/2}]$.  In addition to foreground
absorption, this model requires three parameters: a 1 GHz radio flux
$S_9$, a mean radio-to-X-ray spectral index $\alpha$ ($S_\nu \propto
\nu^{-\alpha}$), and a rolloff frequency $\nu_{\rm roll}$, the
``critical frequency'' for electrons with energy $E_{\rm max}$,
related to $E_{\rm max}$ by $E_{\rm max} = 39 (h \nu_{\rm roll}/1 \
{\rm keV})^{1/2} (B / 10\ \mu{\rm G})^{1/2}$ TeV.  In applying this
model in Paper I, we used $S_9 = 0.9$ Jy, and obtained $\nu_{\rm roll}
= 1.4 \times 10^{18}$ Hz, but with a very high absorbing column
density, $N_H = (5.5 \pm 0.3) \times 10^{22}$ cm$^{-2}$, implying
significant scattering by dust along the line of sight.  Such
scattering removes photons from the source and distributes them in a
faint halo out to arcminute distances, but can also redistribute
photons across the source, in both cases in an energy-dependent
fashion.

Below we present a reanalysis of the {\it Chandra} observation using
Markov chain Monte Carlo (MCMC) techniques, including the effects of
dust scattering, with an emphasis on characterizing spatial
variations.  Such variations hold important information on how the
acceleration of electrons and/or magnetic-field amplification depends
on different conditions such as shock speed and the obliquity angle
\tbn\ between the shock normal and upstream magnetic field,
information crucial in understanding shock acceleration in different
astrophysical environments.

\section{Spectral Analysis Methods}

Radial averages of {\it Chandra} data show a dust-scattered halo out
to about $3'$.  Thus, the spectral analysis of G1.9+0.3 must be
conducted jointly with a halo analysis.  First, we refit the
integrated spectrum without scattering, using a different abundance
set for absorption \citep{grsa98}, and a slightly different, improved
background model.  We then performed a joint analysis of the
background-subtracted spectra from the source (innermost ellipse) and
halo (between two outer ellipses) regions shown in
Figure~\ref{haloimspec}.

We consider a simple uniform dust distribution along the line of sight
to G1.9+0.3, using total and differential X-ray scattering cross
sections from \citet{draine03}.  Small-angle scattering by dust out of
the G1.9+0.3 source extraction region attenuates the X-ray spectrum by
a factor of $\exp (-\tau_{sca}) + f_{src}(1-\exp (-\tau_{sca}) )$,
where $\tau_{sca}$ is the energy-dependent optical depth for
scattering, and $f_{src}$ is the energy-dependent fraction of photons
that have been scattered into the source extraction region. Unless
there are large quantities of dust in the immediate vicinity of the
SNR, it is sufficient to consider only singly scattered photons in the
second term containing $f_{src}$. The fraction of singly scattered
photons is equal to $\tau_{sca}\exp (-\tau_{sca})$
\citep[e.g,][]{malee91}, so scattering attenuates the source spectrum
by $(1+f_{src}\tau_{sca})\exp (-\tau_{sca})$.  The fraction $f_{src}$
depends on the assumed dust distribution and the spatial structure of
the X-ray emitting gas. For a point source, the angular distribution
of singly-scattered photons can be readily found from eq.~(19) of
\citet{draine03}.  We approximated the spatial structure of G1.9+0.3
by the platelet-smoothed (Willett 2007) image shown in
Figure~\ref{xrim}, and convolved it with the model point source halo
to arrive at $f_{src}$. Similarly, the halo region flux is
proportional to $f_{halo}(1-\exp (-\tau_{sca}))$, or
$f_{halo}\tau_{sca}\exp (-\tau_{sca})$ for singly scattered
photons. We neglect multiply scattered photons in modeling of the
G1.9+0.3 dust halo -- this underpredicts the halo intensity at low
energies where $\tau_{sca}$ exceeds unity, by at most 30\% and
probably less (Smith et al.~2006).  We investigated the effect of the
{\it Chandra} point-spread function (PSF), but found that only about
1\% of the SNR flux at high energies is scattered within the telescope
into the halo extraction annulus.

We performed joint modeling of the G1.9+0.3 and halo spectra within
the XSPEC software package \citep{arnaud96}, with dust scattering and
PSF effects implemented as local multiplicative models. Model input
parameters are energy-dependent functions $f_{psf}$ (the
energy-dependent fraction of the SNR flux scattered into the halo
extraction region), $f_{src}$, and $f_{halo}$, and a scalar parameter
$N_H^{sca}$ that provides scaling for $\tau_{sca}$ in terms of the
effective H column density for dust scattering.  Under a
single-scattering approximation, $f_{src}$ and $f_{halo}$ depend only
on the dust distribution along the line of sight, not on $\tau_{sca}$,
so they may be specified prior to modeling spectra. We use the {\tt
srcut} model described above for the intrinsic X-ray spectrum.  While
background subtracted spectra are shown for clarity in Figure
\ref{haloimspec}, we do not subtract background prior to modeling so
that we may use MCMC methods as implemented in XSPEC and
PyMC\footnote{Patil, Huard, \& Fonnesbeck (2009) PyMC: Markov chain
Monte Carlo for Python, version 2.0. Available at
http://pymc.googlecode.com.}. We determined the normalization
parameter $S_9$ from our 1.5 GHz radio image, extrapolated to 1 GHz
using a spectral index of 0.62 \citep{green08}, obtaining a value of
1.17 Jy (under the possibly incorrect assumption that all the radio
flux comes from the same population of electrons as produces the
X-rays).  The absorbing and scattering column densities may differ if
any of the assumptions underlying our uniform-dust model break down,
e.g., an excess of dust local to the SNR will not affect absorption
but will reduce large-angle scattering, resulting in $N_H >
N_H^{sca}$.

It is important to model the background correctly in view of the
faintness of the dust scattered halo. We extracted a background
spectrum from a large ($7.6' \times 3.2'$ in size) region on the ACIS
S3 chip beyond the dust-scattered halo. To separate sky and particle
backgrounds, we also extracted spectra from the most recent stowed
(particle only) ACIS S3 background file available online at the
CXC\footnote{http://cxc.harvard.edu/contrib/maxim/acisbg}. This
particle-only spectrum was fit with a combination of Gaussians and
power laws, exponentially cut off on both ends. We then modeled our
background spectrum as a combination of a sky background, consisting
of two absorbed power laws, and a particle-only background (allowing
for change in intensity only). This fit, after appropriate scaling and
again allowing for variations in the particle background intensity,
was used in the modeling reported below.

The Bayesian MCMC methods require specification of priors on unknown
(fitted) parameters. Uniform, noninformative priors were used for
$N_H$, $N_H^{sca}$, $\alpha$, $\log \nu_{\rm roll}$, $\Gamma$, and
$\log F_X$, with the following exceptions. In the spatially-resolved 
five-region analysis, we employed a normal prior for $\alpha$
with the measured radio index of 0.62 and estimated standard (1
$\sigma$) error of 0.02. A normal prior for $N_H$ was assumed for
protruding SE and NW extensions (``ears''), based on results of that
fit. Since results are prior-dependent in this case, we also
imposed an upper cutoff of $\nu_{\rm roll} = 10^{21}$ Hz on the
corresponding prior, as theoretical considerations make higher rolloff
frequencies unlikely.

\section{Modeling and Results}

We quantified the azimuthal brightness variations obvious in
Figure~\ref{xrim} by circumscribing the shell with an elliptical
annulus of width $19''$, and dividing it into 18 equal segments.  We
found about a factor of 6 variation between the bright NW and SE limbs
and the fainter regions between.

Table~\ref{fitsc} shows results for modeling the integrated spectrum
with the {\tt srcut} model.  Column 1 shows the improved no-scattering
parameter values (means derived from MCMC chains) and their 5 and 95
percentile scores.  Column 2 shows the corresponding values with our
simple scattering model.  The chief difference is a lowering of \roll\
by about 50\%.  The new value, $5.4 \times 10^{17}$ Hz ($h\nu_{\rm
roll} = 2.2$ keV) is still among the highest for a shell SNR. In the
third and fourth columns we give results from a two-zone model, with
the zones comprising most of the bright emission from the SE and NW
limbs, in addition to a halo.  Absorptions were tied together for the
two regions, while the contribution from scattering from the
``missing'' faint limbs and center was accounted for by artificially
increasing scattering by 20\%, required to obtain consistency in the
derived $N_H$ values with the value from the spatially-integrated
model.  The spectral indices $\alpha$ differ at the 99.98\% level, and
\roll\ values at the 94\% level.  A similar simulation neglecting dust
scattering gave qualitatively similar results but higher \roll\ values
by over a factor of 2, indicating the importance of including dust
scattering.  These regions include about 80\% of the X-ray counts, but
less than half the radio, suggesting that some of the radio flux may
not be associated with the blast wave.

\begin{figure}
\vspace{-0.2truein}
\epsscale{1.35}
\plotone{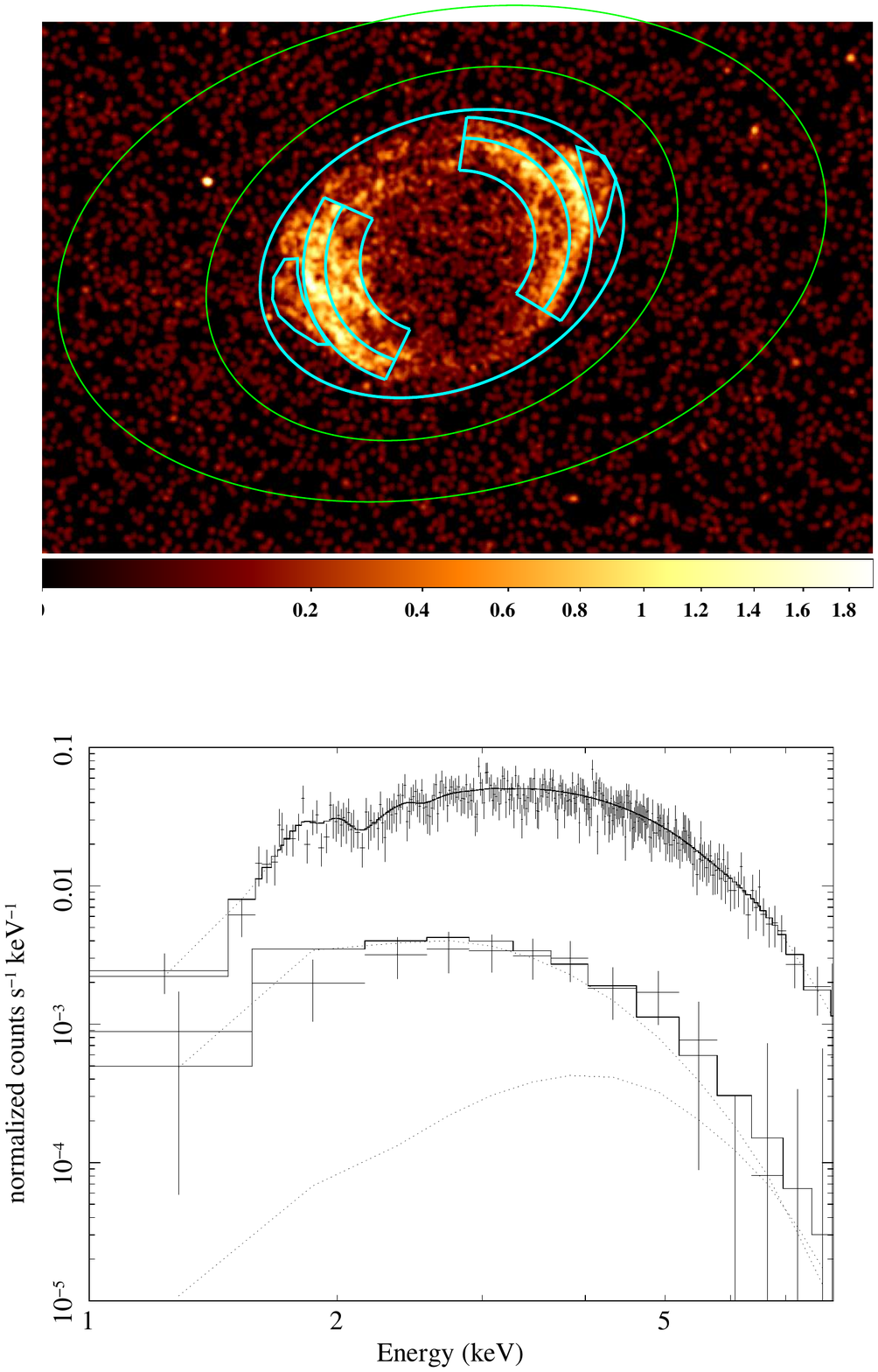}
\caption{Top: Raw {\sl Chandra} image of \src, showing source (cyan)
and background (green) regions.  The six regions used for
spatially-resolved modeling are indicated (the ``ears'' are the sum of
the two irregular regions, and the faint limbs and center comprise the
remainder of the source).  Colors are intensities only (ct
pixel$^{-1}$), between 1.5 and 6 keV.  The image has been smoothed
with a Gaussian with $\sigma = 3$ pixels ($1.5''$).  Bottom:
Integrated spectrum of \src\ and of its halo (top and bottom,
respectively), extracted from regions shown above, after background
subtraction and adaptive binning (both done for display purposes
only).  Model: {\tt srcut} modified by scattering (see text). For the
halo spectrum, contributions from scattering by dust and within the
telescope are shown separately (upper and lower dotted curves,
respectively).  Two point sources within the dust-scattering halo
region (between outermost two ellipses) were excluded.
\vspace{0.1truein}
\label{haloimspec}}
\end{figure}

Spectral results at higher angular resolution are desirable.  We
subdivided \src\ into six regions as shown in Figure~\ref{haloimspec},
but the X-ray data were not adequate to permit the use of the
scattering model.  Since scattering moves photons both out of the
source and from brighter regions into fainter ones, the predominant
effect is to smooth gradients, in both total brightness and spectrum,
so these results should underestimate the true spatial variations.
For a model-independent description of the observed data, we described
the six regions simultaneously with power laws, tying the absorption
columns together, but we also performed {\tt srcut} modeling.  Here
the clearly discrepant radio properties of the ``ears'' made it
necessary to impose a different prior on $\alpha$ (flat) than for the
five regions within the source ellipse, so the ``ears'' modeling was
performed separately.  We also bounded the flat prior on log \roll\ by
log $\nu_{\rm roll}({\rm max}) = 21.0$.  Table~\ref{fit6} gives both
power-law photon indices $\Gamma$ ($N(h\nu) \propto (h\nu)^{-\Gamma}$
ph cm$^{-2}$ s$^{-1}$) and {\tt srcut} parameters.

The systematic trends with brightness are apparent: the opposing
bright limbs of the shell have harder spectra (either lower $\Gamma$
or higher {\roll}) than the center, and the spectra soften also moving
to the interior.  The projecting fainter ``ears'' beyond the E and W
peaks were combined for better statistics, and have the flattest
spectra of all.  The correlation of X-ray brightness with \roll\
strongly suggests a single cause.  Our primary observational result is
this systematic variation of spectral hardness, correlated with
brightness.  A complete dust-scattering model, including the effects
of scattering back into source regions, applied to the spatially
resolved data will require better X-ray data, but while the numerical
values for \roll\ will change somewhat, we do not expect these
trends to be reversed.

\section{Discussion}

We summarize our results as follows:

\begin{enumerate}

\item Including dust scattering lowers \roll\ by about 50\% in the
integrated spectrum, and by factors of more than 2 in the two-region
fit, significantly lowering the inferred maximum electron energies.
However, the \roll\ values are still among the highest ever reported.
The mean spectrum between radio and X-rays is significantly flatter in
the SE than in radio alone for our two-zone model.

\item Both X-ray brightness and spectral flatness or \roll\ show
strong bilateral symmetry.  As one moves away from the bright SE and
NW limbs in azimuth or toward the center, the mean surface brightness
drops by at least 6, while values of \roll\ drop by about an order of
magnitude.

\item The radio structure is dominated by a bright shell with somewhat
smaller peak radius than the X-ray shell, and a single intensity maximum.

\item The ``ears'' extending beyond the SE and NW rims in X-rays show the 
hardest spectra of all, and are not detected at radio wavelengths.

\end{enumerate}

The relatively poor correlation of radio and X-ray brightness around
the shell, and the smaller radius of the peak of the radio shell,
suggest that much of the radio emission may not originate at the outer
blast wave, but further inside, perhaps due to electron acceleration
and/or turbulent magnetic-field amplification at the contact
discontinuity between shocked ISM and shocked ejecta, as is probably
the case in Cas A (e.g., Cowsik \& Sarkar 1984).  We focus on the
X-ray structure and spectra.

In principle, the systematic azimuthal brightness variations could be
due to an inhomogeneous upstream medium that has the bilateral
symmetry of G1.9+0.3.  This explanation requires higher densities in
the SE and NW, where they would be expected to impede the expansion.
However, the remnant has a larger SE-NW than NE-SW diameter.  It is
equally unlikely to imagine systematically larger magnetic fields on
opposite sides of the remnant.  However, the observed simple bilateral
symmetry of both brightness and spectral variations can arise
naturally if electron injection, acceleration rate, or magnetic-field
amplification depend somehow on shock obliquity {\tbn}.  (The physical
origin of such dependences is not clear.)  If the magnetic field is
roughly uniform and not too far from the plane of the sky, then as a
spherical shock encounters it, the obliquity angle around the remnant
periphery will vary with the observed bilateral symmetry.  That
explanation essentially requires the supernova to be Type Ia, as a
core-collapse supernova is expected to expand into stellar-wind
material with a frozen-in magnetic field in a tightly wound Parker
spiral (roughly azimuthally oriented) with all obliquities therefore
near $90^\circ$.

Synchrotron brightness from a power-law distribution of electrons
$N(E) = K E^{-(2\alpha + 1)}$ varies as $K B^{1 + \alpha}$.  The
electron energy density can vary with obliquity if the injection or
acceleration rate depends on \tbn, while the magnetic-field strength
will vary due to simple flux-freezing and shock compression of the
tangential component, and possibly also to turbulent
amplification due to cosmic-ray-induced instabilities
\citep[e.g.,][]{bell01}, a process which might also have obliquity
dependence.  However, in all known cases, X-ray synchrotron emission
is produced by electrons on the loss-steepened tail of a power-law
distribution \citep[e.g.,][]{reynolds99}, introducing a third
parameter.  For the simple {\tt srcut} model, that parameter is the
rolloff frequency $\nu_{\rm roll}.$ X-ray brightness will vary even
for constant $K$ and $B$ if \roll\ varies, and brighter regions will
have harder spectra.

For a uniform upstream magnetic field, if the post-shock field is
simply compressed, a strong-shock compression ratio of 4 will produce
post-shock magnetic fields four times larger where the shock is
perpendicular (\tbn $\sim 90^\circ$), and hence a larger synchrotron
emissivity by $4^{1 + \alpha} = 9.5$ compared to where \tbn $\sim
0^\circ$ (using $\alpha = 0.62$; Green et al.~2008).  This is adequate
to explain the required factor of about 6, but would imply that the
bright limbs are projections of an equatorial ``belt'' which might be
detectable across the remnant center in a deeper observation.

Standard first-order Fermi shock acceleration in the test-particle
limit predicts a spectral index $\alpha = 0.5$ for a strong shock.
Steeper radio indices are universal in young remnants, perhaps
indicating cosmic-ray-modified shocks (e.g., Reynolds \& Ellison
1992).  We find a steeper radio-to-X-ray value of $\alpha$ than that
in the radio alone, consistent with the expectation of modified
shocks.  Better radio data will be required to confirm the radio index
and search for possible spatial variations.

In Fermi shock acceleration, three possible limitations on $E_{\rm
max}$ are synchrotron losses ($E_{m1}$), finite age (or size;
$E_{m2}$), and escape due to an abrupt change in upstream diffusion
properties ($E_{m3}$)(Reynolds 1998).  These depend differently on
physical parameters, including shock obliquity, which may affect the
acceleration rates \citep{jokipii87}.  If we lump all such obliquity
dependence into a factor $f_{tbn}(\theta_{\rm Bn})$, the corresponding
rolloff frequencies obey $\nu_{\rm roll} \propto E_{\rm max}^2 B$ and
we have
\begin{equation}
\nu_{m1} \propto f_{tbn} \,u_8^2 \qquad 
  \nu_{m2} \propto f_{tbn}^2 \, B_2^3 \, t^2 \, u_8^4 \qquad 
  \nu_{m3} \propto \lambda_m^3 \, B_1^2 \, B_2
\end{equation} 
where $B_1$ and $B_2$ are the upstream and downstream magnetic-field
strengths, respectively, $u_8 \equiv u_{\rm sh}/10^8$ cm s$^{-1},$ and
$\lambda_m$ is a wavelength of MHD waves above which the diffusion
coefficient is assumed to jump to much larger values, mimicking the
absence of longer waves capable of scattering more energetic
particles.

The absence of $B$-dependence in $\nu_{m1}$ means that if electron
energies are limited by radiative losses, the different spectral
slopes cannot be produced purely by varying the magnetic field.  The
strong shock-speed dependence could produce a significant effect; but
the observed order-of-magnitude variation in \roll\ would require a
greater variation in $u_{\rm sh}$ than seems consistent with the
remnant's circular outline to within about 10\%.

Since $B_2/B_1$ is obliquity-dependent, either age-limited (case 2) or
escape-limited (case 3) acceleration can explain some spectral
variations with obliquity even if $f_{tbn} \equiv 1$.  A non-linear
variant of escape-limited acceleration in which $\lambda_m$ grows as
increasingly higher-energy particles escape might be able to provide
the observed range; otherwise one expects $\lambda_m$ roughly
constant, a property of the upstream medium, giving a maximum
variation of $\nu_{m3}$ of 4 or so.  Age-limited acceleration could
easily accomplish this without additional obliquity-dependence, since
$B_2^3$ will vary by a factor of at least 64.  We conclude that a
minimal explanation of both brightness and spectral variations is
interaction of the blast wave with a roughly uniform magnetic field
parallel to the bright rims (roughly SW -- NE), with no additional
obliquity-dependence of either electron injection or of acceleration
rates.  In that case, the ``ears'' must represent enhanced upstream
diffusion where the shock is perpendicular, which seems unlikely since
cross-field diffusion mean free paths are probably limited to the
gyroradius. However, a preshock-diffusion explanation is consistent
with the lack of radio emission from the ``ears'' (as lower-energy
radio-emitting electrons would not diffuse as far ahead of the shock).
In this case, an {\tt srcut} model for the ``ears'' would not be
appropriate, as already suggested by their discrepant value of
$\alpha$.

An alternative explanation involves a magnetic field in which the
shock is parallel at the ``ears.''  This would almost certainly
require enough turbulent amplification of magnetic field to swamp the
frozen-in field increase where the shock is perpendicular.  In
addition, the magnetic field would need to be quite close to the plane
of the sky, so that the bright limbs, now ``polar caps'' seen edge-on,
are at the edge of the remnant in projection.  The same requirement
occurs for this geometry in SN 1006.  Strong magnetic-field
amplification probably requires efficient shock acceleration, so that
a population of energetic ions is implied.  However, for any
magnetic-field geometry, effects due solely to magnetic-field
dependence of acceleration time cannot produce the spectral variations
if electron acceleration is limited by synchrotron losses.

Currently scheduled observations may allow considerable clarification.
Our 250-ks {\sl Chandra} follow-up, and the higher-resolution VLA
study we have undertaken, should allow better understanding of the
relation of radio to X-ray electron populations, better
characterization of the scattering, and better deduction of spatially
resolved spectral properties.  \src\ will provide valuable constraints
on the obliquity-dependence of shock acceleration and magnetic-field
amplification.

\acknowledgments

This work was supported by NASA through Chandra General Observer
Program grant GO6-7059X.


\begin{deluxetable}{lcccc}
\tablecolumns{5}
\tablewidth{0pc}
\tabletypesize{\footnotesize}
\tablecaption{Spectral Fits}

\tablehead{
\colhead{Region} & Whole     & Whole  & \multicolumn{2}{c} {Two regions (scat)}\\
   \colhead{}        & (no scat)& (scat) & SE & NW}

\startdata
$N_H$ (abs) ($\times 10^{22}$ cm$^{-2}$) & 6.76 (6.37, 7.16)       & 5.06 (4.48, 5.62)    & \multicolumn{2}{c}{5.12 (4.56, 5.68)} \\

$N_H$ (scat)                            &  ---                 & 3.48 (2.68, 4.35)    & \multicolumn{2}{c}{3.60 (2.83, 4.41)}  \\

$S_9$ (Jy)   & 1.17                 & 1.17                 & 0.203 & 0.291\\

$\alpha$     & 0.649 (0.625, 0.673)    & 0.634 (0.614, 0.655) &0.566 (0.541, 0.592) & 0.612(0.588, 0.637)\\

log \roll\ (Hz) & 17.90 (17.58, 18.29) & 17.73 (17.48, 18.01) &17.48 (17.23, 17.76) & 17.72 (17.44, 18.06)\\

\enddata

\tablecomments{Errors are 90\% confidence limits throughout.  The no-scattering
fit updates the values from Paper I.}
\label{fitsc}
\end{deluxetable}

\begin{deluxetable}{lccccc}

\tablecolumns{6} 
\tablewidth{0pc} 
\tabletypesize{\footnotesize}
\tablecaption{Multi-Region Spectral Parameters} 

\tablehead{ \colhead{Region} &$S_9$ (mJy) & $\alpha$ & log $\nu_{\rm roll}$ (Hz) 
  & $\Gamma$ & Flux}

\startdata
Outer SE     &  67 & $0.60 \pm 0.02$ & 18.12 (17.80, 18.52) & 2.2 (2.1, 2.4) & 6.5 (6.2, 6.9)\\ 
Inner SE     & 190 & $0.62 \pm 0.02$ & 17.46 (17.24, 17.70) & 2.6 (2.4, 2.8) & 4.4 (4.2, 4.7)\\
Faint limbs, 
  center     & 600 & $0.63 \pm 0.02$ & 17.16 (16.98, 17.35) & 2.8 (2.6, 3.0) & 5.8 (5.4, 6.2)\\
Outer NW     &  68 & $0.63 \pm 0.02$ & 18.55 (18.06, 19.21) & 2.1 (1.9, 2.3) & 4.8 (4.5, 5.1)\\
Inner NW     & 220 & $0.64 \pm 0.03$ & 17.69 (17.43, 18.00) & 2.4 (2.2, 2.6) & 4.7 (4.4, 5.0)\\
``Ears''  &$<2$ & $<0.51 \pm 0.04$ & $<$ 18.67 (17.65, 20.37) & 2.0 (1.7, 2.2) & 3.4 (3.1, 3.7)\\

\enddata

\tablecomments{Errors are 90\% confidence limits.  Fitted absorption
$N_H = 7.3 \ (6.8, 7.7) \times 10^{22}$ cm$^{-2}$ for power-law models
and $6.9 \ (6.6, 7.1) \times 10^{22}$ cm$^{-2}$ for {\tt srcut}
models.  Flux is in units of $10^{-13}$ ergs cm$^{-2}$ s$^{-1}$
between 1 and 8 keV. ``Ears'' {\tt srcut} modeling was performed
independently: we assumed a flat prior on \roll\ to a maximum log
$\nu_{\rm roll} = 21.0$, and a flat prior on $\alpha$.}

\label{fit6}
\end{deluxetable}


\newpage

\begin{thebibliography}{}

\bibitem[Arnaud(1996)]{arnaud96}
Arnaud, K.A. 1996, in Astronomical Data
Analysis Software and Systems V, eds. G.~Jacoby \& J. Barnes
ASP Conf.~Ser.~101, 17

\bibitem[Bell \& Lucek(2001)]{bell01}
Bell, A.R., \& Lucek, S.G.
2001, MNRAS, 321, 433

\bibitem[Cowsik \& Sarkar(1984)]{cowsik84}
Cowsik, R., \& Sarkar, S.
1984, MNRAS, 207, 745

\bibitem[Draine(2003)]{draine03}
Draine, B.~T. 2003, ApJ, 598, 1026

\bibitem[Green et al.(2008)]{green08}
Green, D.A., et al.
2008, MNRAS, 387, L54

\bibitem[Grevesse \& Sauval(1998)]{grsa98}
Grevesse, N., \& Sauval, A. J.
1998, Space Science Reviews, 85, 161

\bibitem[Jokipii(1987)]{jokipii87}
Jokipii, J.R.
1987, ApJ, 313, 842

\bibitem[Mathis \& Lee(1991)]{malee91}
Mathis, J. S., \& Lee, C. W.
1991, ApJ,  376, 490

\bibitem[Reynolds(1998)]{reynolds98}
Reynolds, S.P. 1998, ApJ, 493, 375 (R98)

\bibitem[Reynolds(2008)]{reynolds08a}
Reynolds, S.P. 2008, ARA\&A, 46, 89

\bibitem[Reynolds et al.(2008)]{reynolds08b}
Reynolds, S.P., et al.
2008, ApJ, 680, L41 (Paper I)

\bibitem[Reynolds \& Ellison(1992)]{reynolds92}
Reynolds, S.P., \& Ellison, D.C.
1992, ApJ, 399, L75

\bibitem[Reynolds \& Keohane(1999)]{reynolds99}
Reynolds, S.P., \& Keohane, J.W.
1999, ApJ, 525, 368

\bibitem[Smith et al.(2006)]{smith06}
Smith, R.K., et al.
2006, ApJ, 648, 452


\bibitem[Willett(2007)]{willett07}
Willett, R. 2007, in Statistical Challenges in
Modern Astronomy IV, eds.~G.J.~Babu \& E.D.~Feigelson, 
APS Conf.~Ser.~371, 247


\end{thebibliography}
\end{document}